\documentclass[11pt,reqno,addw]{article}
\usepackage[cp1251]{inputenc}
\usepackage[russian]{babel}

\begin{document}


\begin{center}
{\bf \large{Wave Function of Particle and Coordinates Distribution
in Relativistic Quantum Theory}}
\vspace*{1.5mm}

V. F.~ Krotov\\
\vspace*{1.5mm} {\em Institute of Control Problems, Russian
Academy of Sciences, Russia, 117997, Moscow, Profso'uzna'a 65
}\\
\end{center}
\vspace*{1mm}

{\scriptsize{The conditions for observation of the individual
particle coordinates, required by logic of the Special Relativity
and filtering the quantum field effects, are described. A general
relation between the corresponding density of probability and the
wave function is found. It is a relativistic invariant describing
probability of the particle emergences in space–time. This density
is concretized for bosons, both scalar and vector (including
photon), charged and neutral, also electron. The Heisen\-berg’s
uncertainty relations have been approved in regards to
relativistic particle. As applied to the quantum field, this new
construction is transformed to new characteristic of the particles
distributi\-on in space-time, which complete distribution
throughout impulses. The operators of these distribu\-tions and
the invariant relativistic description for free quantum fields
have been obtained. These new properties of the particles and
fields are proposed for experimental investigations. }}
\vspace*{5mm}

{\flushleft{PACS  03+p; 03.65-w; 03.65 Ta}}

\section*{1. Introduction}

\hspace{0.5cm} Model of the Quantum Particle (QP) combines the
deterministic dynamics of Wave Function (WF) and the statistical
relation of WF with observed values: the dynamic part of QP’s
description and the statistical one. While the former one is quite
well formalized, the second one for relativistic QP (RQP) causes
problems: no formal constructs with coordinates distribution
properties, and the restrictions on observation possibilities
caused by effects of the Quantum Field (QF) are available. These
problems have caused a general opinion that  ``in consistent
relativistic quantum mechanics the particle’s coordinates at all
cannot serve as dynamic variables, $\ldots$ and WF, as information
carriers, cannot be present in it'' \cite{bib1}, page 16. We have
to add: relations of uncertainty of Heisenberg are devoid of the
theoretical rationale as applied to RQP, as there is no law of
coordinate’s distribution necessary for that. We demonstrate that
both problems mentioned above are solved when the statistical part
of RQP have been included in logic of the Special Relativity (SR)
sufficiently fully. Despite the common opinion, there appears a
theoretical possibility to preserve information properties of the
WF in the extent comparable to non-relativistic quantum mechanics
(QM), and there are observation procedures, filtering effects of
QF. This inclusion require correction of the full set of the
observables and of the observation procedures. A correction of the
requirements of covariance follow from this.  Formal constructs
expressed via WF are obtained, that possess all necessary
properties of probabilistic distributions of the observed values,
including particle coordinates. In simplest cases this expression
is formally similar to probability density of the coordinates of
the non-relativisic particle, but has different content. Other
observation conditions, other transformational properties, other
meaning: probability density of the events in space-time, instead
of the probability density of positions in space. Instead of
stochastic dance of the particle, described by the flow of
probability, we have probabilistic distribution of the particle
appearance in space-time , which can not be, generally,
represented by the motion in space. On this basis relations of
uncertainty obtain the justification as applied to RQP. As applied
to the quantum field, this new construct is transformed to new
characteristic of the particles distribution in space-time, which
complete distribution throughout impulses. The operators of these
distributions and the invariant relativistic description for free
quantum fields have been obtained.

\section*{2. Нерелятивистская частица}

\hspace{0.5cm} Since the statistical part is sufficiently
formalized only for the non-relativistic QM, let us start here
with defining its necessary details limited to spinless QP. Let
$\mathbf{x}=\{x^i\}=(x^1,x^2,x^3)$ be the vector of particle’s
spatial coordinates,  the element of the related Euclidian space
$\mathbf{E}$. Complex WF  $\psi (\mathbf{x})$ is defined on the
$\mathbf{E}$, and is considered as the element of the Hilbert
space $\mathbf{H}$  with the norm $L_2(\mathbf{E})$ and respective
product $(\psi_1,\psi_2)={\displaystyle\int\limits_
\mathbf{E}}\psi_1(x)\psi^*_2(x)d\mathbf{E}$, where $d\mathbf{E}$
is elementary  volume of $\mathbf{E}$. As a rule, the WF is finite
in a cube $\mathbf{V}\subset \mathbf{E}$:  $\psi (\mathbf{x})=0$,
$\mathbf{x}\notin \mathbf{V}$, and this  integral is defined.The
evolution of the WF $\psi (t,\mathbf{x})$ on the interval $(0,T)$
describes the QP movement. It satisfies Schrodinger equation,
whereas $\|\psi (t,\mathbf{x})\|$ is its dynamic invariant,
normalized by the condition $\|\psi (t,\mathbf{x})\|=1$.

{\em On observation procedure}. Full set of characteristics with
values $y=\linebreak =\{y_i\}$ is being observed. They are
either components of the vector  $\mathbf{x}$, or the  energy,
impulse, and momentum. Let the operations of realization of QP
movement during the time $T$ be defined in identical physical
conditions, corresponding to given WF evolution $\psi
(t,\mathbf{x})$. At a given moment of time $t$ a single session of
observation (session) is being performed over each realization.
Values $y$ in each session are defined, and are random would they
repeat. Observation of trajectories $y(t)$ is also admissible in
cases when they are identical to the time series of results of
sessions of the independent realizations. For each fixed function
$\psi (t,\mathbf{x})$ and each $t$ the averages  $\langle y\rangle
\left[\psi (t,\mathbf{x})\right]=\{\langle y_i\rangle (\psi )\}$ are defined,
corresponding to infinite set of the sessions.

{\em On formal description}. Given fixed $t$ the averages $\langle
y\rangle \left[\psi (t,\mathbf{x})\right]=\{\langle y_i\rangle
(\psi )\}$ are Hermit’s Quadratic Forms (HQF) in $\mathbf{H}$. If
$y=\mathbf{x}$, then the existence of the probability density
$f(t,\mathbf{x})$ and its relation to WF $f(t,\mathbf{x})=|\psi
(t,\mathbf{x})|^2$ are postulated. Correspondingly, the probability
measure $P(Q/t)=P[Q,\psi (t,\mathbf{x})]=\linebreak
={\displaystyle\int\limits_\mathbf{E}}f(t,x)d\mathbf{E}$ is
defined. If $y=\{y_i\}$ is an aggregate of energy, impulse, and
momentum, than the equality $\langle y\rangle =J(\psi )$   is
postulated, where the right side is the aggregate of related
dynamic invariants of the Wave Equation (WE), HQF $J_i(\psi
)=(\psi ,L_i\psi )$, and $L_i$ is a respective operator (operator
of the\linebreak $i$-th observable). In this case the family of
distributions $P(Q,\psi )$ is expressed through components of
spectral functions of the operators $L_i$ in a known manner.

There exist exceptions from the rule $\psi \in L_2(\mathbf{E})$,
when the norm is not defined. In this case the condition
$P(\mathbf{E},\psi)=1$ does not hold. But relative probabilities
$P(Q_1,\psi)/P(Q_2,\psi)$ are defined. Let us note that $\psi
(\mathbf{x})\in L_2(\mathbf{V})$ in any fixed cube
$\mathbf{V}\subset \mathbf{E}$, and WF is the element of Hilbert
space $\mathbf{H}(\mathbf{V})$ with this norm. By fixing here
$\|\psi \|=1$, one can give $P(Q,\psi)$, $Q\subset \mathbf{V}$,
the meaning probability measure  of the positions $\mathbf{x}$,
and $f(t,\mathbf{x})$~--- its density, with the following
amendment to the observation procedure: only the sessions
registering $\mathbf{x}\in \mathbf{V}$ are considered.

\section*{3. Relativistic Particle. General Model Description}

\hspace{0.5cm} Let $E$ be real pseudo-Euclidian space with
coordinate vector $x=\linebreak =\{x^{\alpha}\}=(x^0,\mathbf{x})$,
$\mathbf{x}\in \mathbf{E}$; $x^0=ct$, $t$ is time, $c$ is speed of
light; metric tensor $e=\{e_{\alpha \beta}\}$ is diagonal:
$e_{00}=-1$, $e_{ii}=1$, $i>0$. On a finite bar $V=(0,cT)\times
\mathbf{V}\subset E$ the WF $\psi (x)$   is defined, that maps $V$
to an Euclidian space $U$ with elements $u$ and product $u_1u_2$.
It satisfies the respective wave equation (WE). The WF is
considered as an element of the Hil-bert space $H(V)$ with the
norm $\|\psi(x)\|^2={\displaystyle\int\limits_V}u^2dE$,
$u=\psi(x)$, $dE=dx^0d\mathbf{E}$ (let us denote norm of the space
$\mathbf{H}$ as $\|\psi(t,\mathbf{x})\|$), and respective product
$(\psi_1,\psi_2)$. For each WF the identical physical conditions
are defined, in which the RQP is observed.

{\em Observation procedure} reproduces it for non-relativistic QP
with the following differences. Session is not caused by moment of
time $t$. Thus, time is excluded from the observation process as a
parameter, but is present as a part of argument $x$,  and,
possible, of the observed value $y$. Session lasts during the time
$T$. Other conditions will hold good. The main condition among
them is preservation of uniqueness of $y$ value at session of
measurement in new conditions. We will discuss this condition
below in application to the concrete observable values.  A measure
$P(Q,\psi)$ is defined for each WF, and respectively, the average
value $\langle y\rangle [\psi(x)]$.

{\em Formal description} reproduces  the same for non-relativistic
QP with the following differences. HQF's $P(Q,\psi )$, $J_i(\psi
)=(\psi ,L_i\psi )$, and respectively, the operators of the
observables are defined in $H$, but not in $\mathbf{H}$.
Characteristics $Y$, and thus,  their average $J(\psi )$, possess
relativistic transformational properties. Therefore, $P(Q,\psi )$
is relativistic invariant.  The latter statement follows from
equality $J(\psi )=\langle y\rangle
(\psi)={\displaystyle\int}ydP(\psi )$, since $J(\psi )$ and $y$
have the same tensor dimension. The dimension of probability
density $g(y,\psi)$ is defined by the equality $dP=gdY=inv$, where
$dY$ is elementary volume of the space of observable values.
Detailed elaboration of model with reference to concrete
observable values and types of particles follows below.

\section*{4. Coordinates Observation}

\hspace{0.5cm} {\em On observation procedure}. The full set of the
observed is vector $y=x$ of spatial-time coordinates of a
particle; the result of each session is the fixation of the event
that a particle has appeared in the point $x$ of the space-time
$E$. A content of the session: let the measuring device be an
electronic microscope; a bunch  of electrons probes  the space,
and at some point in time there happens a collision with a
particle-object; such collision, generally, distorts the
observation conditions, corresponding to the given WF (the least
case is the object-photon, that disappears as a result of the
collision); because of this, only this first collision has to be
taken into consideration; the position  of the particle is
identified with the point on the screen,~--- a trace of the single
scattered electron. The measurement should also record the moment
of collision for complete fixation of the event. Only the
sessions, that demonstrate events $x\in V$  has to be taken into
consideration. The causes of the distortion of this condition can
be the following: the session demonstrate an event $x\notin V$,
the collision is not the only, effects of the QF there are.
Admissibility of the latter is not evident, because QP and QF are
different systems. But it is valid (see below, item 6.4).

{\em On the formal description}. The function
\begin{equation}\label{1b1}
g(x)=\psi^2(x),\quad \|\psi (x)\|^2=1
\end{equation}
posses all necessary properties of the probability density of the
events $x\in V$. This density is a relativistic invariant
(elementary volume $dE$ is  invariant). In simplest cases this
expression is formally similar to probability density of the
coordinates of the non-relativistic particle, but has different
content. Other observation conditions, other transformational
properties, other meaning of the function $\psi^2(x)$: probability
density of the events in space-time, instead of the probability
density  of positions in space. The change of meaning corresponds
to the logic of SR and predicts a new property of the RQP. Instead
of stochastic dance of the particle, described by the flow of
probability, we have probabilistic distribution of the particle
appearance in space-time , which can not be, generally,
represented by the motion in space.

Knowing $g(x)$, one can determine density $g_1(\mathbf{x})$ of
probability of positions $\mathbf{x}$, $g_0(x^0)$ of time $x^0$,
density of spatial coordinates at a fixed moment
$g_1(\mathbf{x}/x^0)$:
$$
g_1(\mathbf{x})=\int\limits_{(0,\infty)}g(x)dx^0,\quad
g_0(x^0)=\int\limits_\mathbf{E} g(x)d\mathbf{E},\quad
g_1(\mathbf{x}/x^0)=g(x)/g_0(x^0).
$$

Accordingly to it, we can proof distribution (\ref{1b1}) without
observing time but observing $g_1(x)$.

{\bf 4.1. Scalar Boson}. Space $U$ is one-dimensional, real or
complex, density: $g(x,\psi)=|\psi(x)|^2$, $\|\psi (x)\|=1$, in a
limited beam $V$.
\medskip

{\bf 4.2. Vector Boson}. WF of such particles is the vector
$u(x)$, mapping $E$ into $E$, or into its complex analog $E^*$. We
have: $u^2=\mathbf{u}^2-(u^0)^2$. Euclidean space $U$ is separated
from $E$ by  condition: $u^2\geq 0$. For the latter, it is
necessary and sufficiently that $u^{\prime 0}=0$ in a fixed frame
of reference $x'$. Thus, space $U$ is defined accurate to
transformation $u'\to u$, i.e. to vector $\mathbf{v}$ of speed of
the system $x$ relatively to $x'$. For massive boson $x'$ is the
system of coordinates related to the particle. Therefore, vector
$\mathbf{v}$ is fixed, and space $U$  is Euclidean section of the
pseudo-Euclidean space $E$. Density: $g(x,\psi )=u^2(x)$.

{\em  Massless vector boson}, {\em photon}, has not the system of
coordinates related to the particle, but the lack of the system
$x'$,  $u^{\prime 0}=0$ is not follow from this. Moreover, if this
system is available, then it is not the only, contrary to massive
particle. Really, let's consider a flat wave package such that its
wave vectors are parallel to a vector $\mathbf{k}$. Including
Lorentz condition in the set of field equations  provides:
$u^{\prime 1}=u^{\prime 0}=0$. Assuming $\mathbf{v}\uparrow
\uparrow \mathbf{k}$, similar to massive boson, it is easy to make
sure that density is defined: $u^0=u^1=u^{\prime 1}=u^{\prime
0}=0$, and space $U$ is a plane with basis $u^2$, $u^3$,
independently from $|\mathbf{v}|$. I.\,e., calibration $u^0=u^1=0$
is invariant in subgroup $\mathbf{v}\uparrow \uparrow \mathbf{k}$
of the Lorentz group. It seems natural the following Postulate of
photon: the system $x'$, $u^{\prime 0}=0$ is available. It
determines the density $g(x)=u^2(x)$ and consistently minimizes
distinction of bosons properties: the system of coordinates
related to the particle in this case is absent, but its property
$u^{\prime 0}=0$ is kept  in any system, moving in parallel to
wave vectors. But it is obtained with destruction of the principle
of gradient invariance of electrodynamics. The latter is confirmed
by its experience. But all of it deals with values and
distributions of tensions, energy, impulses, the moments and does
not concern distributions of the photons coordinates. Only an
experiment can determine alternative choice: either this principle
is unapplicable here and (\ref{1b1}) is valid, or (\ref{1b1}) is
not valid
\medskip

{\bf 4.3. Electron}. $U$ is Euclidean space  with elements,
$u=\{u^{\alpha}\}$,  $\alpha =\linebreak =1,2,3,4$, have
transformational properties of the spinor, and  $(u)^2$ is time
component of the vector. WF $u(x)$ is usually considered as a
trajectory in Hilbert space $\mathbf{H}(\mathbf{V})$ with norm
$L_2(\mathbf{V})$:
$$
\|u(t,\mathbf{x})\|^2={\displaystyle\int}u^2(t,\mathbf{x})d\mathbf{E};
\quad u^2=\sum_{\alpha} |u^{\alpha}|^2.
$$
It satisfies the Dirac's  WE, and $\|u(t,\mathbf{x})\|$ is its
dynamic invariant. These properties give the bases for equation:
$u^2(x)=g(\mathbf{x}/x^0)$, $\|u(t,\mathbf{x})\|^2=1$,
\cite{bib2}. Let us introduce a new WF $\psi(x)$, such that
$g(x)=g_0(x^0)u^2(x)=\psi^2(x)$, $\|\psi(x)\|^2=1$. In virtue of
WE and typical boundary conditions it coincides with $u(x)$
accurate to normalization. We have: $g_0(x^0)={\rm const}=1/cT$,
$\psi=\linebreak =(cT)^{-1/2}u$, $\|\psi(x)\|^2=1$. While $T\to
\infty$, $V\to E$ the limit $\|\psi(x)\|^2$ is defined here, if
similar integral over $\mathbf{E}$ is defined. Densities $g(x)$,
$g_1(\mathbf{x}/x^0)$ coincides  accurate to normalization. They
can be constructed both observing $\mathbf{x}$, with parameter
$x^0$, and observing directly $x$.

\section*{5. Observation of energy-impulse}

\hspace{0.5cm} Let us begin consideration with real scalar boson,
WF of which is defined in a finite beam $V\subset E$.
Observable value $y$ is a vector of 4-impulse
$p=\linebreak =(p^0,\mathbf{p})\in E$. Its average is a dynamic invariant,
quadratic form in $H$. Their eigen WF form a family:
$\psi_k=a_k\exp (ip_kx/\hbar )$ with parameters $a_k>0$, $p_k\in
E$, is known discrete row, $p^2_k=-(mc)^2$, values $a_k$ are
defined by normalization ``particle in unit volume'', and
distribution is described in terms of average quantities of
particles $n_k$ with given 4-impulse $p_k$ as a prototype of  QF.
More precisely, $n_k$ is an average number of the measurement
sessions with result $p_k$. In respective space $l_2$ of
coefficients $C=\{C_k\}$ of decomposition
$\psi={\displaystyle\sum\limits_k}C_k\psi_k$: $n_k=|C_k|^2$. And
under an additional condition $\|C\|=1$ it is the unconditional
distribution of probability of a individual particle occurrences
in space of the impulses, the relativistic invariant. The
traditional distributions coincide with the latter accurate to
normalization, but they are attributed with the sense of
conditional distribution at the moment of time $t$. This sense
contradicts their relativistic invariance, and moreover, generates
the known contradiction, \cite{bib1}: it requires instant fixing
of an impulse at measurements, whereas restrictions of accuracy of
 RQP observation require a long fixing. In case of complex WF
the charge is added to energy and  impulse, and in multicomponent
case spin is added. Let us emphasize, that presence of negative
frequencies in decomposition of WF and, accordingly, observation
of a (individual)  particle with different values of charge in
fixed pair of sessions, is admissible, but  not pair occurrence
(such results are not taken into account). It does not contradict
the laws of conservation which should be carried out only on the
average. But it can be limited by external for QM laws, us  law of
the electric charge conservation.

\section*{6. Quantum Field .  The occupation of space-time}

\hspace{0.5cm} New properties of RQP are transformed as applied to
the QF in the form of characteristics of the particles
distributions into the space-time. A suitable  foundation for
this:  the Dirac’s  and Jordan’s conception of QF as  a system of
identical  particles, \cite{bib2}. Let's consider a system of the
N identical particles with common wave functions $\psi(x)\subset
H(V)$. Let $y=\{y_k\}$ be  set of  the QP observables and $\langle
y\rangle =\{Y_k\}$ is this of the system. Let $\{\psi_i\}$ be the
eigen basis of some physical quantity; $n=\{n_i\}$ be the set of
its occupation numbers, $n_i=0,1,2,\ldots ,N$; $\Phi(n)$ be the
symmetrized (respectively, antisymmetrized) WF of the system
expressed in terms of $n$.

{\em On observation procedure}. The operations of the system
realization in identical physical conditions corresponding to
given WF have been defined. A  single session of observation  is
being performed over each realization during the time $T$. In each
session appears, in general, not simultaneously, $N$  particles.
In doing so, for every  particle is fixed value y. Aggregate
characteristics of the system are expressed directly through these
values. For each WF their averages $\langle Y\rangle (\Phi)$
corresponding to infinite set of sessions are defined. The time is
excluded from the process of observation as a parameter.

{\em Formal description}. WF $\Phi(n)$ maps  a set of values $n$
onto complex Euclidean space $\Upsilon$ with the elements $\gamma$
and product $\gamma_1\gamma_2$ and is considered as an element of
a Hilbert space with the product $(\Phi_1,\Phi_2)$, normalized:
$\|\Phi(n)\|^2=1$. The averages $\langle Y\rangle (\Phi)$ are the
HQF: $\langle Y\rangle (\Phi)=(\Phi,\Lambda\Phi)$, $\Lambda$~---
corresponding operators. In particular,
$P(Q,\Phi)=\left(\Phi(n),f(n)\Phi(n)\right)$, where $f(n)=1$,
$n\in \linebreak \in Q$, $f(n)=0$, $n\notin Q$, is probability of the event $n\in
Q$. Second quantization reproduces nonrelativistic analog,
including, in addition to $\Phi(n)$, operators of the
disappearance and the birth of particle $a_i$, $a^+_i$ and they
attributed to a point $x$,~--- wave operators (WO):
\begin{equation}\label{2b1}
\Psi(x)=\sum_i\psi_i(x)a_i,\quad \Psi^+(x)=\sum_i\psi^*_i(x)a^+_i;
\end{equation}
with the follow differences: WO as functions of $x$ are defined in
$H$, but not in $\mathbf{H}$, and time is included in them
symmetrically with spatial coordinates; formalism must be
relativistically invariant; if $\{\psi_i\}$   is a collection of
the plane waves, then this basis is ``doubled'' in virtue of the
appearing states with negative frequencies and, accordingly,~---
additional feature of QP,  the charge, and the related components
in (\ref{2b1}) gain an unified view:  $\psi_i(x)a_i=\linebreak
=\psi^+_i(x)b^+_i$, $\psi^+_i(x)a^+_i= \psi_i(x)b_i$, where $b_i$,
$b_i^+$,   are operators of the appearance and the birth of these
particles. Also the synthesis technique of the operators $\Lambda
=\{\Lambda_k\}$ of the system characteristics $Y=\{Y_k\}$ is
reproduced, including a rule: we record the average for an
individual particle, and produce the replacement:
\begin{equation}\label{3b1}
\begin{array}{c}
\langle y\rangle (\psi^*,\psi)=(\psi
,L\psi)={\displaystyle\int}\psi^*(x)L\psi
(x)dE;\quad \psi \to \Psi(x),\\[2mm]
\psi^*\to \Psi^+(x),\quad
\Lambda=\langle y\rangle \left(\Psi^+(x),\Psi(x)\right) .
\end{array}
\end{equation}
In this product WO are considered as the elements of $H$.
Respectively: $\langle Y\rangle (\Phi)=(\Phi,\Lambda\Phi)$. Let
$\{\psi_i\}$, be the eigen basis of the observed, and $P_i$ be
probabilities of the corresponding  values $y_i$ when  an
individual RQP is observed (or corresponding average numbers of
measurements). Then  without using WO and (\ref{3b1}) and making
replacement $P_i\to n_i$, we may write down
$$
\Lambda =\sum_i n_iy_i;\quad \langle Y\rangle (\Phi)=\biggl(\Phi
,\sum_i y_in_i\Phi\biggr)=\sum_iy_i \langle n_i\rangle ;
$$
$\langle n_i\rangle$~--- the average occupation numbers.

Operators characterizing the coordinates distribution of particles
are lacking in RQM, as for the individual  RQP, and (\ref{1b1}),
(\ref{3b1}) fill this gap.

Operator $\Lambda(Q)$ of the particles amount in a domain
$Q\subset E$ is the follow:
\begin{equation}\label{4b1}
\Lambda(Q)=\left(\Psi^+(x),f(x)\Psi(x)\right)=\int\limits_Q\Psi^+(x)
\Psi(x)
dE,
\end{equation}
where $f(x)=1$, $x\in Q$,  $f(x)=0$,  $x\notin Q$. Operator
$\Lambda(\mathbf{Q})$ of the particles amount in a domain
$\mathbf{Q}\subset \mathbf{E}$ coincides with operator
$\Lambda(Q)$, $Q=\mathbf{Q}\times (0,T)\subset E$. The average
ocupation number of  the domain $Q:\langle N\rangle
(Q,\Phi)=\left(\Phi,\Lambda(Q)\Phi\right)$.

Let $S(\psi^*,\psi)=(\psi,L_S\psi)$ be action functional of QP,
and $S(\Phi^*,\Phi)=\linebreak =(\Phi,\Lambda_s\Phi)$~--- of the
system.  Varying the latter with respect to $\Phi^*(n)$, we get the following  WE:
\begin{equation}\label{5b1}
\Lambda_s\Phi(n)=0,\quad
\Lambda_s=\left(\Psi^+(x),L_s\Psi(x)\right) .
\end{equation}

Let now the number $N$  be not fixed but varies from session to
session. This is consistent with the model of QF in the corpuscular
concept with an accuracy of the non-observed characteristics of
the vacuum state. WF should now be  symmetrized also in $N$,
\cite{bib2}. And the rule (\ref{2b1}), and respectively the
concrete representations of operators $\Lambda$, including
$\Lambda(Q)$, remain valid as well as WE.
\medskip

{\bf 6.1. Relativistic invariance of the field description}. The
number of particles in a given state is a  result of observation,
which does not depend on the choice of coordinates. Accordingly, a
set of occupation numbers is relativistic invariant, as well as
operations on them $a_i$,  $a^+_i$. Therefore, WO have
relativistic transformation properties of the particle WF $\psi$,
and operators $\Lambda$ have  properties of their analogues $L$.
Furthermore,
$$
\begin{array}{c}
\Lambda=\left(\Psi^+(x),L\Psi(x)\right)={\displaystyle\sum\limits_{ij}}
l_{ij}a^+_ia_j, l_{ij}=(\psi_i,L\psi_j); \\[2mm]
\langle y\rangle (\Phi)=(\Phi,\Lambda\Phi)
={\displaystyle\sum\limits_{ij}}l_{ij}(\Phi,a^+_ia_j\Phi).
\end{array}
$$
But $(\Phi,\Phi)$ is invariant, as a corresponding value of the
probability measure, as well as operators $a^+_ia_j$. Thus,
$(\Phi,a^+_ia_j\Phi)$ are invariants too, and the form
$(\Phi,\Lambda\Phi)$ possesses the relativistic  transformation
properties of the form $(\psi ,L\psi)$. This description is fully
invariant,  unlike decomposition of field into system of the
oscillators, which is invariant in general, but contains
noninvariant fragments.  Also this decomposition is not suitable
to description of the coordinates distributions.
\medskip

{\bf 6.2. Representation of the Quantum Field Characteristics in
Terms of Occupation Numbers of the Impulse States}. Let
$\{\psi_i(x)\}$ be the eigen basis of impulse, of spin and of
charge. With regard to energy, impulse, spin, charge (\ref{3b1})
provides the textbook operators. Operator $\Lambda(Q)$ of the
particle amount in the domain $Q\subset E$ was given with
(\ref{4b1}), where basis $\{\psi_i(x)\}$ in WO should be
renormalized: $\|\psi_i(x)\|^2=1$ instead of  ``particle in the
unit volume''. Let write (\ref{4b1}) for the concrete particles.
\medskip

{\em Scalar neutral boson}.  $a_i=b_i$, $\Psi^+(x)=\Psi(x)$;
$\Lambda(Q)=(1/2)\int\limits_Q \Psi^2(x)dE$.

{\em Photon}. In framework of model of  item  4.2:
$\Psi^+(x)=\Psi(x)=\linebreak =\left(\Psi_2(x),\Psi_3(x)\right)$;
$\Lambda(Q)=(1/2){\displaystyle\int\limits_Q}\Psi^2(x)dE$. Here
the WO $\Psi_2(x)$, $\Psi_3(x)$ correspond to the components of
basis $u^2$, $u^3$.
\medskip

{\em The scalar charged boson}.
$\Psi(x)={\displaystyle\sum\limits_i}a_i\psi_i(x)+b^+_i\psi^*_i(x)$,
$\Psi^+(x)=\linebreak
={\displaystyle\sum\limits_i}a^+_i\psi^*_i(x)+b_i\psi_i(x)$;
$\Lambda(Q)={\displaystyle\int\limits_Q} \Psi^+(x)\Psi(x)dE$.
\medskip

{\bf 6.3. Representation of the Quantum Field Characteristics in
Terms of Occupation Numbers of the Space~--- time Cells}. Let
introduce into consideration  eigen  basis of coordinates. For
unification with discrete basis of the  impulse we make it at a
prelimit level of Riemann integral  sums. We split the beam $V$
into collection of beams $v(\xi)$ having volumes $w(\xi)$, every
of them marked with belonging to  one  value $x=\xi$. Define  a
set of functions $\psi(x,\xi)$  with $\xi$ parameter:
$\psi^2(x,\xi)=w^{-2}(\xi)$, $x\in v(\xi)$; $\psi (x,\xi)=0$,
$x\notin v(\xi)$. Approximate $\psi (x)$ with  step-function
$\psi'(x)=\psi (\xi)$, $x\in v(\xi)$. Functions $\psi (x,\xi)$,
$\psi'(x)$ are the elements of finite-dimensional subspace
$H'\subset H(V)$ with orthogonal basis $\psi (x,\xi)$ with
accuracy  of approximation:
\begin{equation}\label{6b1}
\begin{array}{c}
\left(\psi'_1(x),\psi'_2(x)\right)={\displaystyle\sum\limits_{\xi}}
\psi^{\prime *}_1(\xi)\psi'_2(\xi)w(\xi);\quad \|\psi'(x)\|^2=1,\\[2mm]
\|\psi (x,\xi)\|^2=w(\xi)^{-1};\\[2mm]
\psi'(x)={\displaystyle\sum\limits_{\xi}}\psi (\xi)\psi
(x,\xi)w(\xi)\to \psi(x)={\displaystyle\int}\psi (\xi)\delta(x-\xi)
d\xi ,\\[2mm]
w(\xi)\to 0.\\[2mm]
P(Q,\psi'(x))={\displaystyle\int\limits_{Q'}}\psi^{\prime
2}(x)dE={\displaystyle\sum\limits_{\xi}}P(\xi),\
P(\xi)=\psi^2(\xi)w(\xi).
\end{array}
\end{equation}
Here $Q’$ is the minimal set of beams $v(\xi)$ covering $Q$.
Accordingly, complete the second quantization apparatus to the
eigen  basis of coordinates $\psi (x,\xi)$: $n=\{n(\xi)\}$~--- the
set of occupation numbers of the beams $v(\xi)$,
$n(\xi)=\linebreak =0,1,2,\ldots ,N$. Making the replacement $P(\xi)\to
n(\xi)$, we obtain the operator $\Lambda'(Q)$ of the particles
amount in the domain $Q'\subset E$:
$$
\Lambda (Q')= {\displaystyle\int\limits_{Q'}}\Psi^{\prime
+}(x)\Psi'(x)dE={\displaystyle\sum\limits_{\xi}}n(\xi),\ \xi :
v(\xi)\in Q';
$$
$n(\xi)$ are acting as eigen values of operator $\Lambda'(Q)$. The
average amount of particles in $Q'$:
$$
\langle N\rangle
(Q',\Phi)=\left(\Phi(n),{\displaystyle\sum\limits_{\xi}}n(\xi)
\Phi(n)\right)={\displaystyle\sum\limits_{\xi}} \langle
n(\xi)\rangle ,\quad \xi :v(\xi)\in Q';
$$
where $\langle n(\xi)\rangle$ is average particles amount in
$v(\xi)$.
\medskip

{\bf 6.4. An Individual Particle as a  Quantum Field subsystem}.
Let consider an individual particle  with WF $\psi(x)$ in terms of
the second quantization as subsystem $N=1$ of the  system ``QF''.
We define it in the following way: we observe a QF, and  only the
sessions of QF, that bring out $N=1$. We have: $n =(0,0,\ldots
0,1,0, \ldots )$, $n_i = 0,1$. The term ``identical particles''
loses its meaning, and WF $\Phi(n)$ does not contain the
permutation operator. We find an average amount of particles in
domain $Q\subset E:\langle N\rangle
(Q,\Phi)=\left(\Phi,\Lambda'(Q)\Phi\right)$, using the finite
dimensional approximation given above. $\{\psi(x,\xi)\}$~--- is
the eigen basis of the observed. To every $n$ corresponds the
event  $x(n)\in E$. Identify the point $x(n)$ with the mark $\xi$
of the beam $v(\xi)$. We have:  WF $\Phi(n)=\psi(x(n))$ map the
set of all $n$ into $U$. In agreement with (\ref{6b1}) should be
put
$(\Phi_1,\Phi_2)={\displaystyle\sum\limits_n}\Phi_1(n)\Phi_2(n)w(\xi=x(n))$.
We have:
$$\begin{array}{c}
\|\Phi(n)\|^2=1; \langle N\rangle
(Q,\Phi)=\left(\Phi,\Lambda'(Q)\Phi\right)={\displaystyle\sum\limits_n}
\Phi^2(n)=\\[2mm]
{\displaystyle\sum\limits_{\xi}}
\psi^2(\xi)w(\xi)=P(Q',\psi)\to{\displaystyle\int\limits_Q}
\psi^2(\xi)d\xi, w(\xi)\to 0; \\[2mm]
n:x(n)\in Q, \xi:v(\xi)\in
Q'.
\end{array}
$$

Thus, this function $\Phi(n)$ meets the definition of WF, and
observing such subsystem of QF gives the probability density
(\ref{1b1}) of the individual particle.

\section*{7. About Relations of Uncertainties and Observation Accuracy Estimates}

\hspace{0.5cm} In non-relativistic QM the Heisenberg relations of
uncertainties are the effects of the statistical postulates
described above. Within the framework of traditional model of
relativistic QP they no longer have this theoretical basis due to
the absence of the required law of the coordinates distribution.
Nevertheless, they are used in the same way, strictly speaking,
now as an independent postulate. In the model considered these
relations obtain substantiation. Therewith, while in
non-relativistic QM the relation coordinate-impulse and
time-energy are deduced in different ways and have different
sense, \cite{bib3} стр.~ 185~--~188, but herewith, they possess
full formal and semantic symmetry. Theoretical  bottom threshold
of uncertainty of coordinates: $\triangle x>\triangle
x_{\min}=\hbar c/\varepsilon$ ($\varepsilon$~--- energy), caused
by these relations and absence of the QF effects  (for photon this
is an order of its wave length), loses validity, because under the
observation in framework of our procedure QF effects are
admissible.  This theoretical bottom threshold is replaced with
the minimal practically acceptable value of probability of the
individual particle fixing under the observation of QF.

It is required to give a special interpretation of the
uncertainties relation time-speed-impulse $\mathbf{v}\triangle
t\triangle p>\hbar$, and a derived from it relation of impulse
observation accuracy with duration of an observation session
$\triangle t\triangle p>\hbar /c\triangle t$. First of all, does
the concept ``QP speed'' make sense in the given model, and if
yes, which one?

\section*{8. Material for the experiment}

\hspace{0.5cm} It provided by the new properties of RQP and QF
being predicated here: the probability density formulas for
various types of RQP coordinates and formulas for distributions of
amounts of particles into space-time for QF. The simplest variant
to check the latter: to locate   WF $\Phi$  at eigenbasis of
4-impulse, to calculate the corresponding theoretical $\langle
N\rangle (Q,\Phi)$ and comparing it with directly measured
$\langle N\rangle (Q,\Phi)$. A significant new feature of the
model observation: nonconditionality of the measure from the
moment of time $t$. For a time $T$ the repeated reactions of
particles with instrument of observation are possible, and
respectively~--- nonuniqueness value $y$ when measurement. The
latter should be eliminated in any way. Unlike textbook views, an
individual RQP observation under procedures described are not
burdened with problem of preventing effects of QF. The particular
interest represent the  photon coordinates observation. From one
side, it is ideally provided the uniqueness of the collision with
the particle-device as far as it disappears under the reaction.
From the other side the validity of (\ref{1b1}) for it is
stipulated with the additional postulate, which is alternate to
electrodynamic principle of gradient invariance. And an experiment
will determine the alternative: either the given principle is
invalid here, and the (\ref{1b1}) is valid, or  our postulate of
photon is invalid, and it's coordinate distribution is not
determined.

\section*{9. Conclusions}

\hspace{0.5cm} When the statistical part of RQP has been included
in logic of the SR sufficiently fully, there appears, despite
common opinion, a theoretical possibility to preserve informative
proper-ties of the WF in the extent comparable to non-relativistic
QM. There appears a theoretical op-portunity to observe
spatial-time coordinates of the RQP, and the presentation of the
probability density $g(x)=\psi(x)^2$ satisfying all necessary
requirements gets defined. Here WF $\psi(x)$ maps space-time $E$
into Euclidean space $U$, characteristic for each  type of
particles: boson, real and complex, scalar and vector, including
photon, and electron. In application to photon this formula of the
density is conditioned by an additional assumption, that is
alternative to the electro-dynamic principle of gradient
invariance.  Density $g(x)$ in a simplest case is formally alike
the density of the probability of non-relativistic particle's
coordinates, but has other content: other observation conditions,
other properties, other interpretation of density. Observation
sessions are not caused by moment $t$, i.e. time is excluded from
the observation process as a parameter, but is present in the
argument $x$ and the observable value $y=x$. Density now is
relativistic invariant. Instead of stochastic dance of the
particle, described by the probability flow with spatial density
$f(t,\mathbf{x})=|\psi (t,\mathbf{x})|^2$, we have probabilistic
distribution of particle's appearance in space-time with density
$g(x)$, not resulted, in general, to motion in the space. In a
simplest case the idea of such distribution of coordinates was
considered in \cite{bib4}, and \cite{bib5} is a short version of
this paper. The interpretation of distributions of energy-impulse
of RQP is also clarified here. Relations of uncertainty of
Heisenberg are the yield of the postulates of non-relativistic QM.
In the frame-work of traditional model of RQP they no longer have
this theoretical rationale, as there is no law of coordinates'
distribution, necessary for that. These relations obtain the
justification in the model considered. An observation procedure,
filtering effects of QF, have been proposed and justified.
Respective restrictions of the RQP observation precision, caused
by these effects, are abandoned. As applied to the QF, this new
constructs are transformed to new characteristics of the particles
distribution in space-time, which complete distribution throughout
impulses. The operators of these distributions and the invariant
relativistic description for free QF  have been obtained.

Abbreviations: Quantum Particle~--- QP; Wave Function~--- WF;
quantum mechanics~--- QM;  relativistic QP~--- RQP; Special
Relativity~--- SR; Hermitian Quadratic Form~--- HQF; Wave
Equation~--- WE; Relativistic QM~--- RQM.

\end{document}



\begin{center}
{\bf \large{Волновая функция частицы и распределения координат в
релятивистской квантовой теории}}
\vspace*{1.5mm}

В.Ф.~Кротов\\
\vspace*{1.5mm}

{\em Институт проблем управления Российской академии наук,
Профсоюзная, 65, 117997, Москва, Россия}\\

(Дата: Апрель 24, 2008)
\end{center}
\vspace*{1mm}

{\scriptsize{Описываются условия наблюдения координат единичной
частицы, диктуемые логикой специальной теории относительности и
фильтрующие эффекты квантового поля. Находится общая связь
соответствующей плотности  вероятности с волновой функцией.
Это~--- релятивистский инвариант, описывающий вероятность
появлений частицы в пространстве-времени. Она конкретизируется для
бозона, нейтрального и заряженного, скалярного и векторного,
включая фотон, электрона. Отношения неопределенностей Гейзенберга
получают обоснование применительно к релятивистской частице.
Применительно к квантовому полю эти конструкции трансформируются в
новые характеристики распределения частиц в пространстве-времени,
дополняющие таковые для импульсных состояний. Получены операторы
этих распределений для свободных полей и
релятивистски-инвариантное описание последних. Эти новые свойства
частиц и полей  предлагаются для экспериментального
исследования.}}
\vspace*{5mm}

{\flushleft{PACS  03+p; 03.65-w; 03.65 Ta}}

\section*{1. Введение}

\hspace{0.5cm} В модели квантовой системы, и в частности, частицы
(КЧ), синтезированы детерминированная динамика волновой функции
(ВФ) и статистическая связь последней с наблюдаемыми величинами:
динамическая и статистическая части описания КЧ. Если первая
вполне формализована, то проблемы второй для релятивистской КЧ
(РКЧ) привели к общему мнению, что ``в последовательной
релятивистской квантовой механике координаты частиц вообще не
могут фигурировать в качестве динамических переменных, $\ldots$ а
ВФ как носители информации не могут фигурировать в ее аппарате''
\cite{bib1}, стр.~16. Эти проблемы: отсутствие формальных
конструкций, описывающих распределения координат, и ограниченность
применения модели ввиду эффектов квантового поля (КП). Добавим:
утрачивают теоретическое основание отношения неопределенностей
Гейзенберга, поскольку необходимый для их вывода закон
распределения координат отсутствует. Здесь показывается, что,
вопреки сложившемуся мнению, при достаточно полном включении
статистической части РКЧ в логику специальной теории
относительности (СТО), обе эти проблемы находят решение,
сохраняются информативные свойства ВФ в объеме, сравнимом с
нерелятивистской квантовой механикой (КМ) и имеются процедуры
наблюдения частицы, фильтрующие эффекты квантового поля (КП).
Помимо обычных условий, это включение требует коррекции полного
набора наблюдаемых и процедур наблюдения. Из последнего следует и
коррекция требований ковариантности. Получены выраженные через ВФ
конструкции, обладающие всеми необходимыми свойствами
вероятностных распределений наблюдаемых, включая координаты
частицы. Плотность вероятности последних в простейших случаях формально подобна ее нерелятивистскому аналогу, но имеет иное содержание: иные условия
наблюдения, иные свойства, иной смысл. Сеансы наблюдения не
обусловлены моментом времени, т.\,е. время исключается из процесса
наблюдения в качестве параметра, но присутствует в составе
аргумента ВФ и наблюдаемых; соответственно, плотность теперь~---
релятивистский инвариант; вместо стохастического танца частицы,
описываемого потоком вероятности, имеем вероятностное
распределение появлений частицы в пространстве-времени, не
сводимое, вообще, к движению в пространстве.  На этой базе
подтверждаются для РКЧ отношения неопределенностей. Применительно
к КП эти новые конструкции трансформируются в новые характеристики
распределения частиц в пространстве-времени, дополняющие таковые
для импульсных состояний.

\section*{2. Нерелятивистская частица}

\hspace{0.5cm} Поскольку статистическая часть достаточно
формализована только для нерелятивистской КМ, то начнем с фиксации
необходимых здесь ее деталей, ограничившись бесспиновой КЧ. Пусть
$\mathbf{x}=\{x^i\}=(x^1,x^2,x^3)$~--- вектор пространственных
координат частицы, элемент соответствующего эвклидова пространства
$\mathbf{E}$. Комплексная ВФ $\psi (\mathbf{x})$ определена на
$\mathbf{E}$ и рассматривается как элемент гильбертова
пространства $\mathbf{H}$  с нормой $L_2(\mathbf{E})$ и
соответствующим произведением
$(\psi_1,\psi_2)=\int\limits_\mathbf{E}\psi_1(x)\psi^*_2(x)d\mathbf{E}$,
$d\mathbf{E}$~--- элементарный объем $\mathbf{E}$, $^*$~--- значок
комплексного сопряжения. Как правило, ВФ финитна в некотором кубе
$\mathbf{V}\subset \mathbf{E}$:  $\psi (\mathbf{x})=0$,
$\mathbf{x}\notin \mathbf{V}$, и этот несобственный интеграл
определен элементарно. Эволюция ВФ $\psi (t,\mathbf{x})$ на
интервале $(0,T)$ описывает движение КЧ. Она удовлетворяет
уравнению Шредингера, причем $\|\psi (t,\mathbf{x})\|$~--- его
динамический инвариант, нормированный условием $\|\psi
(t,\mathbf{x})\|=1$.

{\em К процедуре наблюдения}. Полная совокупность наблюдаемых
$y=\linebreak =\{y_i\}$~--- либо вектор  $\mathbf{x}$, либо
импульс. Пусть определены операции реализации движения КЧ в
течение времени $T$ в тождественных физических условиях,
отвечающих данной эволюции ВФ $\psi (t,\mathbf{x})$. Над каждой
реализацией в заданный момент времени $t$ проводится единственный
сеанс наблюдения (замер). Значения $y$ в каждом сеансе определены,
а при их повторении~--- случайны. Допустимо и наблюдение
траекторий $y(t)$ в случаях, когда они отождествимы с временным
рядом результатов замеров независимых реализаций. Для каждой
фиксированной функции $\psi (t,\mathbf{x})$ и каждого $t$
определены средние  $\langle y\rangle \left[\psi (t,\mathbf{x})
\right]=\{\langle y_i\rangle (\psi
)\}$, соответствующие бесконечному набору замеров.

{\em К формальному описанию}. Средние суть эрмитовы квадратичные
формы (ЭКФ) в  $\mathbf{H}$:  $\langle y_i\rangle (\psi )=(\psi ,L_i\psi )$,
$L_i$~--- соответствующий оператор (оператор $i$-й наблюдаемой).
Если $y=\mathbf{x}$, то постулируется наличие плотности
вероятности $f(t,\mathbf{x})$  и следующая ее связь с ВФ:
$f(t,\mathbf{x})=|\psi (t,\mathbf{x})|^2$, и соответственно,~---
мера вероятности: $P(Q/t)=P\left[Q,\psi
(t,\mathbf{x})\right]=\linebreak =\int |\psi
(t,\mathbf{x})|^2d\mathbf{E}$, $Q\in \mathbf{E}$. Если $y$~---
импульс, то постулируется равенство $\langle y\rangle [\psi ]=J(\psi)$, где правая часть~--- соответствующий динамический инвариант волнового
уравнения (ВУ), ЭКФ. В этом случае мера вероятности $P(Q,\psi)$ на
подмножествах $Q$ пространства наблюдаемых выражается известным
образом через компоненты спектральных функций операторов $L_i$.

Имеются исключения из правила $\psi \in L_2(\mathbf{E})$, когда
норма не определена и не удовлетворяется условие
$P(\mathbf{E},\psi)=1$. Но определены относительные вероятности
$P(Q_1,\psi)/P(Q_2,\psi)$. Отметим также, что  в любом
фиксированном кубе $\mathbf{V}\subset \mathbf{E}$ ВФ есть элемент
гильбертова пространства $\mathbf{H}(\mathbf{V})$ с нормой
$L_2(\mathbf{V})$. Фиксируя здесь $\|\psi\|=1$, можно сохранить за
$P(Q,\psi)$, $Q\subset \mathbf{V}$, смысл меры вероятности
наблюдения позиций $\mathbf{x}$, а за $f(t,\mathbf{x})$~--- ее
плотности, при следующей поправке к процедуре наблюдения:
учитываются только сеансы, показывающие $\mathbf{x}\in
\mathbf{V}$.

\section*{3. Релятивистская частица. Общее описание модели}

\hspace{0.5cm} Пусть $E$~--- действительное псевдоэвклидово
пространство с координатным вектором
$x=\{x^{\alpha}\}=(x^0,\mathbf{x})$, $\mathbf{x}\in \mathbf{E}$;
$x^0=ct$, $t$~--- время, $c$~--- скорость света; метрический
тензор $e=\{e_{\alpha \beta}\}$~--- диагональный: $e_{00}=-1$,
$e_{ii}=1$, $i>0$. На конечном брусе $V=(0,cT)\times
\mathbf{V}\subset E$ определена ВФ $\psi (x)$, отображающая $V$ в
эвклидово пространство $U$, $m$-мерное, действительное либо
комплексное, c элементами $u$  и произведением
$u_1u_2={\displaystyle\sum\limits^m_1} (u^*)^i_1u^i_2$. Она
удовлетворяет соответствующему ВУ и рассматривается как элемент
гильбертова пространства $H(V)$  с нормой
$\|\psi(x)\|^2={\displaystyle\int\limits_V}u^2dE$, $u=\psi(x)$,
$dE=dx^0d\mathbf{E}$ (норму пространства $\mathbf{H}$ обозначаем
$\|\psi(t,\mathbf{x})\|$) и соответствующим произведением
$(\psi_1,\psi_2)$. Для каждой ВФ определены тождественные
физические условия, в которых наблюдается РКЧ.

{\em Процедура наблюдения} воспроизводит таковую для
нерелятивистской КЧ со следующими отличиями. Замер не обусловлен
моментом времени $t$. То есть время исключается из процесса
наблюдения в качестве параметра, но присутствует в составе
аргумента $x$  и, возможно, наблюдаемой  $y$. Исключение особой
роли времени в  замерах~--- необходимый элемент внедрения логики
СТО. Замер продолжается в течение времени $T$. Остальные условия
остаются в силе. Главное из них: существование и единственность
значения $y$  при замере в новых условиях. Мы обсудим его ниже
применительно к конкретным наблюдаемым. Предполагается, что
определена мера $P(Q,\psi)$  для каждой ВФ, и соответственно,
среднее $\langle y\rangle [\psi(x)]$.

{\em Формальное описание} воспроизводит таковое для
нерелятивистской КЧ со следующими отличиями. ЭКФ
$\langle y_i\rangle (\psi)=(\psi,L_i\psi)$, и соответственно, операторы
наблюдаемых определены в  $H$, а не в  $\mathbf{H}$.
Характеристики  $y$, и соответственно, их средние, обладают
релятивистскими трансформационными свойствами. При этом,
$P(Q,\psi)$~--- релятивистский инвариант. Последнее следует из
равенства  $\langle y\rangle (\psi)=\int ydP(\psi)$, поскольку $\langle y\rangle (\psi)$ и $y$
имеют одинаковую тензорную размерность. Размерность плотности
вероятности $g(y,\psi)$ определяется равенством  $dP=gdY=inv$, где
$dY$~--- элементарный объем пространства наблюдаемых. Далее
следует детализация модели применительно к конкретным наблюдаемым
и типам частиц.

\section*{4. Наблюдение координат}

\hspace{0.5cm} {\em К процедуре наблюдения}. Полный набор
наблюдаемых расширяется до:  $y=x$; результат каждого замера~---
фиксация события: частица появилась в точке $x$
пространства-времени $E$. Содержание замера: пусть прибор~---
электронный микроскоп; пучок электронов ``прощупывает''
пространство, и в некоторый момент происходит столкновение с
частицей-объектом; оно, вообще, нарушает условия наблюдения,
отвечающие данной ВФ (крайний случай~--- объект-фотон, исчезающий
при столкновении), поэтому должно учитываться только одно
столкновение; позиция частицы фиксируется точкой на экране,~---
следом единственного рассеянного электрона. Для полной фиксации
события, необходима также фиксация момента столкновения.
Учитываются только сеансы, показывающие событие  $x\in V$, притом
единственное. Сеансы, нарушающие это условие, исключаются из
рассмотрения. Причины нарушений: показание  $x\notin V$,
неединственность столкновений, эффекты КП. Справедливость
допущения последних и их фильтрации таким способом неочевидна,
поскольку КЧ и КП суть разные системы. Но это так, см.~ниже,
п.~6.4. В этом смысле допускается наблюдение КП.

{\em К формальному описанию}. Выражение
\begin{equation}\label{1b1}
g(x)=\psi^2(x),\quad \|\psi (x)\|^2=1
\end{equation}
обладает всеми необходимыми свойствами плотности вероятности
событий $x\in V$  и предлагается для экспериментальной проверки в
этом качестве. Оно формально подобно его нерелятивистскому
варианту, но имеет иное содержание. Иные условия наблюдения, иные
трансформационные свойства (релятивистский инвариант), иной смысл:
вероятность событий в пространстве-времени вместо позиций в
пространстве. Изменение смысла соответствует логике СТО и
предсказывает новое свойство КЧ. Вместо стохастического танца,
описываемого потоком вероятности, имеем вероятностное
распределение случайных появлений частицы в пространстве-времени,
не сводимое, вообще, к движению в пространстве.

Зная  $g(x)$, можно найти плотность $g_1(\mathbf{x})$  вероятности
позиций $\mathbf{x}$,\linebreak $g_0(x^0)$~--- времени $x^0$,
пространственных координат в фиксированный момент как условную
$g_1(\mathbf{x}/x^0)$:
$$
g_1(\mathbf{x})=\int\limits_{(0,\infty)}g(x)dx^0,\quad
g_0(x^0)=\int\limits_\mathbf{E} g(x)d\mathbf{E},\quad
g_1(\mathbf{x}/x^0)=g(x)/g_0(x^0).
$$
Это значит, что предсказываемые распределения можно проверять, не
наблюдая временную компоненту событий.

{\bf Замечание}. Финитность ВФ  $\psi (x)$  на  $V$, вообще, не
совместна с ВУ, во всяком случае, для бозонов. Это значит, что в
замерах возможны показания  $x\notin V$, также как появление пар и
других эффектов КП. Т.\,е. закон (\ref{1b1}) описывает условное
распределение вероятности на избранном множестве замеров, где
отсутствуют такие результаты.
\medskip

{\bf 4.1. Скалярный бозон}. Пространство $U$  одномерное,
действительное либо комплексное,  плотность:
$g(x,\psi)=|\psi(x)|^2$.
\medskip

{\bf 4.2. Векторный бозон}. ВФ есть вектор  $u(x)$, отображающий $E$  в  $E$, либо в его комплексный аналог   $E^*$. Имеем:  $u^2=\mathbf{u}^2-(u^0)^2$. Эвклидово пространство $U$  выделяется условием:  $u^2\geq 0$. Для последнего необходимо и достаточно: $u^{\prime 0}=0$  в какой–либо фиксированной системе отсчета  $x'$. Пространство $U$  тем самым оказывается определено с точностью до преобразования  $u'\to u$, т.\, е.~--- до вектора скорости  $\mathbf{v}$  системы  $x$ относительно  $x'$. Для {\em массивного бозона} такая система $x'$   существует. Это~--- его система покоя. Тем самым вектор  $\mathbf{v}$  фиксирован и пространство $U$  есть 3 мерное эвклидово сечение псевдоэвклидового пространства  $E$. Плотность:  $g(x,\psi )=u^2(x)$.

{\em  Безмассовый бозон}, {\em фотон}, не имеет системы покоя, но отсюда не следует отсутст-вие системы  $x'$,  $u^{\prime 0}=0$. Более того, в отличие от массивного бозона, если такая система  $x'$  существует, то она не единственна. Действительно, выделим случай, когда поле есть плоский волновой пакет с волновыми векторами компонент, параллельными  вектору  $\mathbf{k}$. Включив условие Лоренца в состав уравнений поля, и  предполагая, что вектор $\mathbf{v}$  параллелен  $\mathbf{k}$, получим:  $u^{\prime 1}=0$. Легко убедиться, что  $u^0=u^1=u^{\prime 1}=u^{\prime 0}=0$, т. е. пространство $U$  есть поперечная плоскость с базисом $u^2$, $u^3$  независимо от  $|\mathbf{v}|$. То есть, калибровка  $u^0=u^1=0$ инвариантна на подгруппе группы Лоренца  $\mathbf{v}\uparrow \uparrow \mathbf{k}$. На основании сказанного представляется естественным доопределить фотон постулатом: существует система  $x'$,  $u^{\prime 0}=0$. Он исключает неопределенность выражения  $g(x)=u^2(x)$  в силу его градиентной неинвариантности и обеспечивает его положительность.  Доопределяется на фотон (\ref{1b1}) и непротиворечиво минимизируется различие свойств бозонов: система покоя для фотона отсутствует, но ее свойство $u^{\prime 0}=0$  сохраняет целое семейство систем. Но достигается это ценой отказа от принципа градиентной инвариантности электродинамики. Последний подтвержден ее опытом (за единственным известным исключением). Но весь он связан со значениями и распределениями напряженностей, энергии, импульсов, моментов и не касается распределения координат фотонов. Только эксперимент может определить альтернативный выбор: либо данный принцип здесь неприменим, и справедливо  $g(x)=u^2(x)$, либо распределение координат фотона не определено.
\medskip

{\bf 4.3. Электрон}. Пространство $U$ 4-мерное, комплексное с
элементами $u=\{u^{\alpha}\}$,  $\alpha =1,2,3,4$,
трансформационные свойства элементов~--- спиноры, а $u^2$~---
временная компонента вектора. ВФ $u(x)=u(x^0,\mathbf{x})$
рассматривается обычно как траектория в гильбертовом пространстве
$\mathbf{H}(\mathbf{V})$. Она удовлетворяет ВУ Дирака, причем
$\|u(x^0,\mathbf{x})\|$~--- его динамический инвариант. Эти
свойства дают основания для равенства: $u^2(x)=\linebreak
=g(\mathbf{x}/x^0)$, $\|u(t,\mathbf{x})\|^2=1$, \cite{bib2}.
Введем новую ВФ $\psi(x)$, такую что $g(x)=\linebreak
=g_0(x^0)u^2(x)=\psi^2(x)$, $\|\psi(x)\|^2=1$. В силу ВУ она
совпадает с $u(x)$ c точностью до нормировки. Имеем:
$g_0(x^0)={\rm const}=1/cT$, $\psi=(cT)^{-1/2}u$,
$\|\psi(x)\|^2=1$. При $T\to \infty$, $V\to E$ предел
$\|\psi(x)\|^2$ здесь определен, если определен аналогичный
несобственный интеграл по  $\mathbf{E}$. Плотности $g(x)$,
$g(\mathbf{x}/x^0)$ совпадают с точностью до нормировки и могут
синтезироваться как путем наблюдения $\mathbf{x}$  с $x^0$ в
качестве параметра, так и непосредственно $x$.

\section*{5. Наблюдение энергии-импульса}

\hspace{0.5cm} Пусть КЧ~---  действительный скалярный бозон.
Наблюдаемая~--- вектор 4-импульса  $p=(p^0,\mathbf{p})\in E$. Его
среднее есть соответствующий динамический инвариант, ЭКФ в  $H$.
Собственные ВФ образуют семейство: $\psi_k=a_k\exp (ip_kx/\hbar )$
с параметрами $a_k>0$, $p_k\in E$~--- известный дискретный ряд,
$p^2_k=-(mc)^2$, значения $a_k$ определяются нормировкой ``частица
в единичном объеме'', и распределение описывается в терминах
средних количеств частиц $n_k$ с данным 4-импульсом $p_k$  в
качестве наглядного полуфабриката квантового поля. А точнее,
$n_k$~--- среднее количество замеров с результатом  $p_k$. В
соответствующем пространстве  $l_2$ коэффициентов $C=\{C_k\}$
разложения $\psi={\displaystyle\sum}_kC_k\psi_k$: $n_k=|C_k|^2$, а при
дополнительном условии $\|C\|=1$ это~---
релятивистски-инвариантное безусловное распределение вероятности
появлений единичной частицы в пространстве 4-импульсов.
Хрестоматийные распределения совпадают с последними с точностью до
нормировки, но им приписывается смысл условного распределения в
момент времени  $t$. Этот смысл противоречит их релятивистской
инвариантности, и кроме того, порождает известное противоречие
\cite{bib1}: он требует мгновенной фиксации импульса при замерах,
тогда как ограничения  точности наблюдения РКЧ требуют
продолжительной фиксации. В случае комплексной ВФ к энергии и
импульсу добавляется заряд, а многокомпонентной~--- спин.
Подчеркнем, что допускается, вообще, и наличие отрицательных
частот в разложении ВФ и, соответственно,~--- наблюдение частицы
(единственной) с разными значениями заряда в фиксированной паре
сеансов, но~--- не появление пары (такие показания не
учитываются). Это не противоречит квантовому закону сохранения
заряда, выполняемому только в среднем, но может не допускаться
внешними для КМ законами, такими, как всеобщий закон сохранения
электрического заряда.

\section*{6. Квантовое поле. Заполнение пространства-вре\-ме\-ни}

\hspace{0.5cm} Новые свойства РКЧ доопределяются на КП в виде
характеристик распределения частиц в пространстве-времени.
Адекватная база для этого: концепция КП как соответствующей
системы тождественных частиц Дирака и Иордана, \cite{bib2}.
Рассмотрим систему $N$ тождественных частиц с одинаковой ВФ
$\psi(x)\subset H(V)$. Пусть $y=\{y_k\}$~--- наблюдаемая
совокупность характеристик одной КЧ и $Y=\{Y_k\}$~--- системы.
Пусть    $\{\psi_i\}$~--- собственный базис некоторой физической
величины;   $n=\{n_i\}$~--- набор его чисел заполнения,
$n_i=0,1,2,\ldots ,N$; $\Phi(n)$~--- симметризованная (либо,
соответственно, альтернизированная) ВФ системы, выраженная через
$n$.

{\em К процедуре наблюдения}. Определены операции реализации
системы в тождественных физических условиях, отвечающих данной ВФ.
Над каждой реализацией проводится единственный сеанс наблюдения
(замер), продолжающийся в течение времени  $T$. В каждом сеансе
появляется, вообще, не одновременно, $N$ частиц. При этом для
каждой частицы фиксируется единственное значение  $y$.
Агрегированные характеристики $Y$  системы выражаются
непосредственно через эти значения. Для каждой ВФ определены их
средние  $\langle Y\rangle (\Phi)$, соответствующие бесконечному набору замеров.
Время исключается из процесса наблюдения в качестве параметра.

{\em Формальное описание}. ВФ $\Phi(n)$ отображает множество
значений $n$ в комплексное эвклидово пространство $\Upsilon$ с
элементами $\gamma$ и произведением $\gamma_1\gamma_2$ и
рассматривается как элемент гильбертова пространства с
произведением  $(\Phi_1,\Phi_2)$, нормированный: $\|\Phi(n)\|^2=1$.
Средние $\langle Y\rangle (\Phi)$ суть ЭКФ: $\langle Y\rangle (\Phi)=(\Phi,\Lambda\Phi)$,
$\Lambda$~--- соответствующие операторы. В частности,
$P(Q,\Phi)=\left(\Phi(n),f(n)\Phi(n)\right)$, где $f(n)=1$, $n\in
Q$, $f(n)=0$, $n\notin Q$, есть вероятность события $n\in Q$.
Вторичное квантование воспроизводит нерелятивистский аналог,
включая, помимо $\Phi(n)$, операторы уничтожения и рождения
частицы $a_i$, $a^+_i$, и таковые, отнесенные к точке $x$,~---
волновые операторы (ВО):
\begin{equation}\label{2b1}
\Psi(x)={\displaystyle\sum}_i\psi_i(x)a_i,\quad
\Psi^+(x)={\displaystyle\sum}_i\psi^*_i(x)a^+_i.
\end{equation}
Отличия следующие: ВО как функции $x$ определены в  $H$, а не в
$\mathbf{H}$, и время входит в них симметрично с пространственными
координатами; формализм должен быть релятивистски инвариантен; если $\psi_i$~---
плоские волны, то этот базис известным образом расширен, соответственно, появляется
дополнительная характеристика КЧ,Ё--- заряд, а соответствующие новым состояниям слагаемые
в (\ref{2b1}) обретают унифицированный вид $\psi_i(x)b_i$, $\psi^*_i(x)b^+_i$, где $b_i$,
$b^+_i$~--- операторы уничтожения и рождения частицы в этих состояниях.
Воспроизводится также техника синтеза операторов $\Lambda
=\{\Lambda_k\}$ характеристик системы   $Y=\{Y_k\}$, включая
правило: записываем среднее для единичной частицы, и производим замену:
\begin{equation}\label{3b1}
\begin{array}{c}
\langle y\rangle (\psi^*,\psi)=(\psi ,L\psi)=\int\limits \psi^*(x)L\psi
(x)dE;\quad \psi \to \Psi(x),\\[2mm]
\psi^*\to \Psi^+(x),\quad
\Lambda=\langle y\rangle \left(\Psi^+(x),\Psi(x)\right) .
\end{array}
\end{equation}
Здесь в произведении ВО рассматриваются как элементы $H$.
Соответственно: $\langle Y\rangle (\Phi)=(\Phi,\Lambda\Phi)$. Если
$\{\psi_i\}$~--- собственный базис наблюдаемой, то $\Lambda
={\displaystyle \sum\limits}_in_iy_i$;  $\langle Y\rangle (\Phi
)=\biggl({\displaystyle
\sum\limits}_iy_in_i\Phi\biggr)={\displaystyle
\sum\limits}_iy_i\langle n_i\rangle $; $y_i$~--- собственные числа
оператора $L$, а $n_iy_i$~--- оператора $\Lambda$, $\langle
n_i\rangle$~--- средние числа заполнения. Операторы,
характеризующие распределения координат частиц, отсутствуют в РКМ,
как и для единичной РКЧ, и (\ref{1b1}), (\ref{3b1}) восполняют
этот пробел. Оператор $\Lambda (Q)$ количества частиц в области
$Q\subset E$:

\begin{equation}\label{4b1}
\Lambda(Q)=\left(\Psi^+(x),f(x)\Psi(x)\right)=\int\limits_Q\Psi^+(x)\Psi(x)
dE,
\end{equation}
где  $f(x)=1$, $x\in Q$,  $f(x)=0$,  $x\notin Q$. Оператор
$\Lambda(\mathbf{Q})$ количества частиц в области
$\mathbf{Q}\subset \mathbf{E}$ совпадает с оператором
$\Lambda(Q)$, $Q=\mathbf{Q}\times (0,T)\subset E$. Среднее число
заполнения области $Q:\langle N\rangle
(Q,\Phi)=\left(\Phi,\Lambda(Q)\Phi\right)$.

Пусть $S(\psi^*,\psi)=(\psi,L_S\psi)$~--- функционал действия КЧ,
а $S(\Phi^*,\Phi)=\linebreak =(\Phi,\Lambda_s\Phi)$~--- системы.
Варьируя последний по $\Phi^*(n)$, получим ВУ:
\begin{equation}\label{5b1}
\Lambda_s\Phi(n)=0,\quad
\Lambda_s=\left(\Psi^+(x),L_s\Psi(x)\right) .
\end{equation}

Пусть теперь число $N$  не фиксировано, а меняется от замера к
замеру. Это соответствует модели КП в рамках корпускулярной
концепции с точностью до ненаблюдаемых характеристик вакуумного
состояния. ВФ  теперь должна быть симметризована еще и по  $N$,
\cite{bib2}. Правило же (\ref{2b1}), и соответственно,~---
конкретные представления операторов  $\Lambda$, включая
$\Lambda(Q)$, остаются справедливыми, также как и ВУ.
\medskip

{\bf 6.1. Релятивистская инвариантность описания поля}. Количество
частиц в данном состоянии есть результат наблюдения, не зависящий
от выбора системы координат. Соответственно, набор чисел
заполнения релятивистски инвариантен, так же как и операции над
ним   $a_i$,  $a^+_i$. Поэтому ВО обладают релятивистскими
трансформационными свойствами ВФ частицы  $\psi$, а операторы
$\Lambda$~--- свойствами их аналогов $L$. Далее,
$\Lambda=\left(\Psi^+(x),L\Psi(x)\right)={\displaystyle\sum}_{ij}l_{ij}a^+_ia_j$,
$l_{ij}=(\psi_i,L\psi_j)$; $\langle y\rangle \langle Y\rangle (\Phi)=\linebreak =
(\Phi,\Lambda\Phi)
={\displaystyle\sum}_{ij}l_{ij}(\Phi,a^+_ia_j\Phi)$. Но $(\Phi,\Phi)$ есть
инвариант как соответствующее значение меры вероятности, также как
и операторы $a^+_ia_j$, поэтому $(\Phi,a^+_ia_j\Phi)$~--- тоже
инварианты, а форма $(\Phi,\Lambda\Phi)$ обладает релятивистскими
трансформационными свойствами формы $(\psi ,L\psi)$. Данное
описние  ``насквозь'' релятивистски инвариантно, в отличие от
декомпозиции поля на осцилляторы, инвариантной в целом, но
содержащей неинвариантные звенья. Не говоря уже о том, что
последняя никак не соотносится с распределениями координат.
\medskip

{\bf 6.2. Представление характеристик квантового поля в числах
заполнения импульсных состояний}. Пусть  $\{\psi_i(x)\}$~---
собственный базис энергии, импульса, спина и заряда. Применительно к энергии,
импульсу, спину, заряду (\ref{3b1}) дает хрестоматийные операторы.
Оператор  $\Lambda(Q)$  количества частиц в области $Q\subset E$
задан (\ref{4b1}), причем базис $\{\psi_i(x)\}$ в ВО должен быть
перенормирован: $\|\psi_i(x)\|^2=1$ вместо ``частицы в единичном
объеме''. Запишем (\ref{4b1}) для конкретных частиц.

{\em Скалярный нейтральный бозон}.  $a_i=b_i$,
$\Psi^+(x)=\Psi(x)$; $\Lambda(Q)=\linebreak =(1/2)\int\limits_Q
\Psi^2(x)dE$.

{\em Фотон}. В рамках модели п.~3.3:
$\Psi^+(x)=\Psi(x)=\left(\Psi_2(x),\Psi_3(x)\right)$;
$\Lambda(Q)=(1/2)\int\limits_Q \Psi^2(x)dE$. Здесь ВО $\Psi_2(x)$,
$\Psi_3(x)$ соответствуют компонентам базиса $u^2$, $u^3$.

{\em Скалярный заряженный бозон}.
$\Psi(x)={\displaystyle\sum}_ia_i\psi_i(x)+b^+_i\psi^*_i(x)$,
$\Psi^+(x)=\linebreak ={\displaystyle\sum}_ia^+_i\psi^*_i(x)+b_i\psi_i(x)$;
$\Lambda(Q)=\int\limits_Q \Psi^+(x)\Psi(x)dE$.
\medskip

{\bf 6.3. Представление характеристик квантового поля в числах
заполнения ячеек пространства-времени}. Введем в рассмотрение
собственный базис координат, для унификации с дискретным базисом импульса~--- на допредельном уровне интегральных сумм. Разобъем брус $V\subset E$ на брусы
$v(\xi)$ с объемом $w(\xi)$, каждый помечен значением $x=\xi$,
принадлежащим ему. Определим семейство функций  $\psi(x,\xi)$  с
параметром $\xi$: $\psi^2(x,\xi)=w^{-2}(\xi)$, $x\in v(\xi)$;
$\psi (x,\xi)=0$, $x\notin v(\xi)$. Аппроксимируем $\psi (x)$
ступенчатой функцией $\psi'(x)=\psi (\xi)$, $x\in v(\xi)$.
Функции $\psi (x,\xi)$, $\psi'(x)$ суть
элементы подпространства $H'\subset H(V)$ конечной размерности с
ортогональным базисом $\psi (x,\xi)$  и, с точностью до аппроксимации:
\begin{equation}\label{6b1}
\begin{array}{c}
\left(\psi'_1(x),\psi'_2(x)\right)={\displaystyle\sum}_{\xi}\psi^{\prime
*}_1(\xi)\psi'_2(\xi)w(\xi);\quad \|\psi'(x)\|^2=1,\\[2mm]
\|\psi (x,\xi)\|^2=w(\xi)^{-1};\\[2mm]
\psi'(x)={\displaystyle\sum}_{\xi}\psi (\xi)\psi (x,\xi)w(\xi)\to
\psi(x)=\int\limits \psi (\xi)\delta(x-\xi)d\xi ,\\ [2mm]
w(\xi)\to 0.\\[2mm]
P(Q,\psi'(x))={\displaystyle\int}_{Q'}\psi^{\prime 2}(x)dE={\displaystyle\sum}_{\xi}P(\xi),\ P(\xi)=\psi^2(\xi)w(\xi).
\end{array}
\end{equation}
Здесь $Q’$~--- минимальный набор брусов  $v(\xi)$,  покрывающий $Q$. Соответственно, доопределим аппарат вторичного квантования на
собственный базис координат   $\psi (x,\xi)$:   $n=\{n(\xi)\}$~---
множество чисел заполнения брусов  $v(\xi)$, $n(\xi)=0,1,2,\ldots
,N$. Производя замену: $P(\xi)\to n(\xi)$, получим оператор количества частиц в области $Q'\subset E$:
$$
\Lambda (Q')={\displaystyle\sum}_\xi n(\xi),\ \xi : v(\xi)\in Q';
$$
$n(\xi)$ выполняют здесь роль собственных чисел  оператора
$\Lambda(Q')$. Среднее количество частиц в $Q'$:
$$
\langle N\rangle
(Q',\Phi)=\left(\Phi(n),{\displaystyle\sum}_{\xi}n(\xi)\Phi(n)\right)={\displaystyle\sum}_{\xi}
\langle n(\xi)\rangle ,\quad \xi :v(\xi)\in Q';
$$
$\langle n(\xi)\rangle$~--- среднее количество частиц в  $v(\xi)$.
\medskip

{\bf 6.4. Единичная частица как подсистема квантового поля}.
Рассмотрим единичную частицу с ВФ $\psi(x)$  в терминах вторичного
квантования как подсистему  $N=1$ системы ``КП'', определяемую
следующим образом: учитываются только замеры КП, показывающие
$N=1$. Имеем: $n = (0,0,\ldots 0,1,0,\ldots )$, $n_i = 0,1$.
Термин ``тождественные частицы'' теряет смысл, и ВФ $\Phi(n)$  не
содержит оператора перестановки. Найдем среднее количество частиц
в области $Q\subset E:\langle N\rangle
(Q,\Phi)=\left(\Phi,\Lambda'(Q)\Phi\right)$, пользуясь
конечномерной аппроксимацией, приведенной выше.
$\{\psi(x,\xi)\}$~--- собственный базис наблюдаемой. Каждому $n$
отвечает событие $x(n)\in E$. Отождествим точку $x(n)$ с пометкой
$\xi$ включающего ее бруса  $v(\xi)$. Имеем: ВФ
$\Phi(n)=\psi(x(n))$ отображает множество всех $n$  в   $U$. В
согласии с (\ref{6b1}) следует положить
$(\Phi_1,\Phi_2)={\displaystyle\sum}_n\Phi_1(n)\Phi_2(n)w(\xi=x(n))$. Имеем:
$\|\Phi(n)\|^2=1$; $\langle N\rangle
(Q,\Phi)=\left(\Phi,\Lambda'(Q)\Phi\right)={\displaystyle\sum}_n\Phi^2(n)={\displaystyle\sum}_{\xi}
\psi^2(\xi)w(\xi)=\linebreak =P(Q',\psi)\to
{\displaystyle\int\limits_Q} \psi^2(\xi)d\xi$, $w(\xi)\to 0$;
$n:x(n)\in Q$, $\xi:v(\xi)\in Q'$. Таким образом, данная функция
$\Phi(n)$  отвечает определению ВФ, а наблюдение такой подсистемы
КП дает тот же результат, что и описанное выше непосредственное
наблюдение единичной частицы.

\section*{7. Об отношениях неопределенностей и оценках точности наблюдения}

\hspace{0.5cm} В нерелятивистской КМ отношения неопределенностей
Гейзенберга суть следствие статистических постулатов, приведенных
выше. В рамках традиционной модели релятивистской КЧ они уже не
имеют этого теоретического основания, поскольку необходимый для
этого закон распределения координат отсутствует. Тем не менее, они
используются в том же виде, строго говоря, уже в качестве
самостоятельного постулата. Здесь эти отношения получают
обоснование. При этом, если в нерелятивистской КМ отношения
координаты-импульс и время-энергия выводятся разным способом и
имеют разный смысл \cite{bib3} стр.~ 185~--~188, то здесь они
обладают полной формальной и смысловой симметрией. Теоретический
нижний порог неопределенности координат РКЧ: $\triangle
x>\triangle x_{\min}=\hbar c/\varepsilon$ ($\varepsilon$~--- энергия),
обусловленный этими отношениями и недопущением эффектов КП (для
фотона,~--- порядка длины его волны), теряет  силу, поскольку
эффекты КП при наблюдении в рамках описанной здесь процедуры
допустимы. Он заменяется минимальным практически
приемлемым значением вероятности фиксации единственной частицы при
наблюдении КП.

Требует специального осмысления отношение неопределенностей
время~--- скорость~--- импульс $\mathbf{v}\triangle t\triangle
p>\hbar$ и вытекающая из него связь точности наблюдения импульса с
продолжительностью сеанса наблюдения   $\triangle p>\hbar
/c\triangle t$. Прежде всего, имеет ли смысл понятие ``скорость
КЧ'' в данной модели, и если да, то какой?

\section*{8. Материал для эксперимента}

\hspace{0.5cm} Его дают новые свойства РКЧ и КП, предсказываемые
здесь: формулы для плотности вероятности координат различных типов
РКЧ и распределений количеств частиц в пространстве-времени для
КП. Простейший вариант проверки последнего:  фиксация ВФ   на
собственном базисе 4-импульса, подсчет соответствующего
теоретического $\langle N\rangle (Q,\Phi)$  и сравнение его с
непосредственно замеренным  $\langle N\rangle (Q,\Phi)$.
Существенная новая особенность модели наблюдения:
необусловленность замера моментом времени  $t$. В течение времени
$T$ возможны повторные реакции частиц с прибором, и
соответственно,~--- неединственность значения $y$  при замере.
Последняя должна быть так или иначе устранена. В отличие
от хрестоматийного мнения, наблюдение единичной РКЧ согласно
описанным процедурам не обременено проблемой недопущения эффектов
КП. Особый интерес представляет наблюдение координат фотона. С одной стороны,
идеально обеспечивается единственность столкновения с частицей-прибором,
поскольку он исчезает при реакции. С другой,~--- применимость к нему (\ref{1b1})
обусловлена дополнительным постулатом, альтернативным принципу градиентной
инвариантности электродинамики. И эксперимент определит альтернативу: либо данный
принцип здесь неприменим, и справедливо (\ref{1b1}), либо несправедлив наш постулат
фотона, и его распределение координат не определено.

\section*{9. Выводы}

\hspace{0.5cm} Показано, что при достаточно полном включении
статистической части РКЧ в логику СТО информативные свойства ее ВФ обретают объем,
сравнимый с нерелятивистской КМ. Появляется, вопреки
хрестоматийному мнению, теоретическая возможность наблюдения
координат РКЧ, и определено представление их плотности вероятности
$g(x)=\psi^2(x)$, отвечающее всем необходимым требованиям. Здесь
ВФ $\psi(x)$ отображает пространство-время $E$ в эвклидово
пространство $U$, характерное для каждого из рассмотренных типов
частиц: бозона, действительного и комплексного, скалярного и
векторного, включая фотон, электрона. Плотность $g(x)$ в простейших случаях
формально подобна плотности вероятности координат нерелятивистской частицы,
но имеет иное содержание: иные условия наблюдения, иные свойства, иной смысл.
Сеансы наблюдения не обусловлены моментом времени $t$, т.\,е. время исключается
из процесса наблюдения в качестве параметра, но присуствует в составе аргумента
ВФ и наблюдаемой $y=x$; соответственно, плотность теперь~--- релятивистский инвариант;
вместо стохастического танца частицы, описываемого потоком вероятности с пространственной
плотностью $f(t,\mathbf{x})=|\psi (t,\mathbf{x})|^2$, имеем вероятностное распределение
появлений частицы в пространстве-времени с плотностью $g(x)$, не сводимое, вообще, к
движению в пространстве. В первом приближении идея такого распределения координат была
рассмотарена в \cite{bib4} на простейшем примере скалярного бозона. Сокращенная версия
данной работы опубликована в \cite{bib5}. Уточняется  также смысл распределений
энергии-импульса РКЧ. Отношения неопределенностей Гейзенберга суть следствие постулатов
нерелятивистской КМ. В рамках традиционной
модели РКЧ они уже не имеют этого теоретического основания,
поскольку необходимый для этого закон распределения координат
отсутствует. В рассматриваемой модели эти отношения получают
обоснование. Предложена и обоснована процедура наблюдения частицы,
фильтрующая эффекты квантового поля КП. Соответственно, теряют
силу ограничения точности наблюдения РКЧ основанные на недопущении
этих эффектов. Новые свойства РКЧ доопределяются на КП в виде характеристик распределения частиц в пространстве-времени в дополнение к таковым для импульсных состояний. Получены операторы этих распределений для некоторых
свободных КП, а также~--- релятивистски инвариантное описание КП в
рамках предлагаемой модели.

Аббревиатуры: КЧ~--- квантовая частица; ВФ~--- волновая функция;
РКЧ~--- релятивистская КЧ; КМ~--- квантовая механика; РКМ~---
релятивистская КМ;  СТО~--- специальная теория относительности;
КП~--- квантовое поле; ЭКФ~--- эрмитова квадратичная форма; ВУ~---
волновое уравнение; ВО~--- волновой оператор.